\documentclass[12pt,reqno]{article}
\usepackage{amsfonts}
\usepackage{amsmath}
\usepackage{amsbsy}

\usepackage{amssymb,latexsym}
\numberwithin{equation}{section}

\begin{document}
 \allowdisplaybreaks[1]
\title{Dualisation of the Salam-Sezgin D=8 Supergravity}
\author{Tekin Dereli\\
Department of Physics,\\
Ko\c{c} University,\\
Rumelifeneri Yolu 34450,\\
Sar$\i$yer, \.{I}stanbul, Turkey.\\
\texttt{tdereli@ku.edu.tr},\\
\\
Nejat T. Y$\i$lmaz\\
          Department of Physics,\\
          Middle East Technical University,\\
          06531 Ankara, Turkey.\\
          \texttt{ntyilmaz@metu.edu.tr}}
\maketitle
\begin{abstract}The first-order formulation of the Salam-Sezgin $D=8$ supergravity
coupled to N vector multiplets is discussed. The non-linear
realization of the bosonic sector of the $D=8$ matter coupled
Salam-Sezgin supergravity is introduced by the dualisation of the
fields and by constructing the Lie superalgebra of the symmetry
group of the doubled field strength.

\end{abstract}

\section{Introduction}

It is possible to formulate the scalar sectors of a wide class of
supergravity theories as non-linear sigma models based on $G/K$
coset parametrization maps. In particular the scalar sectors of
all the pure and the matter coupled $N>2$ extended supergravities
in $D=4,5,6,7,8,9$ dimensions as well as the maximally extended
supergravities in $D\leq11$ can be formulated as symmetric space
sigma models. The first-order bosonic field equations of the
maximal supergravities for $D\leq11$ are regained as a twisted
self-duality equation by the non-linear realization of the bosonic
sector of these theories in [1,2]. Therefore the non-linear nature
of the scalar fields has been modified to cover the bosonic field
content as well. In [3,4,5] the non-linear realization has been
enlarged further to include the graviton for the $D=11$, IIB and
IIA supergravities. In these later works including the gravity the
entire bosonic sector is formulated as a coset model by
constructing the algebra which generates the coset and by taking
the Lorentz group as the local symmetry group. Furthermore in the
same works it is also discussed that a larger non-linear
realization of the bosonic sector can be considered to include a
Kac-Moody algebra identified as $E_{11}$.

In this work we formulate the bosonic sector of the $D=8$,
Salam-Sezgin supergravity which is coupled to N vector multiplets
[6] as a non-linear realization. The 2N vector multiplet scalars
of the theory parameterize the coset $SO(N,2)/SO(N)\times SO(2)$
which is a Riemannian globally symmetric space [7] and $SO(N,2)$
is a semi-simple, non-compact real form while $SO(N)\times SO(2)$
is its maximal compact subgroup. The construction presented here
is parallel with the one of [1,2]. We will define the Lie
superalgebra which will be used to formulate the bosonic fields as
a non-linear coset model. The algebra structure will be
constructed in such a way that it will yield the integrated
first-order bosonic equations of the theory as a twisted
self-duality equation of the Cartan form of the coset
representative [1,2].

Section two is the introduction of the $D=8$, matter coupled
Salam-Sezgin supergravity. We will derive the equations of motion
and then locally integrate them to find the first-order field
equations. Whereas in section three by following the outline of
[1,2] we will introduce dual fields for the bosonic field content
of the theory excluding the graviton and we will also define new
algebra generators for the original fields and their duals. A
coset element will then be constructed whose Cartan form is
intended to yield the correct second-order equations when inserted
in the Cartan-Maurer equation. In order to calculate the Cartan
form one needs to know the algebra structure of the generators
which parameterize  the coset representatives when coupled to the
fields. The algebra which generates the coset representatives is a
differential graded algebra and it is composed of the differential
forms and the field generators. It covers the Lie superalgebra of
the field generators which is the Lie algebra of the symmetry
group of the Cartan form. The correct choice of the commutators
and the anti-commutators of the Lie superalgebra is a result of
the direct comparison of the field equations of section two and
the Cartan-Maurer equation. Finally we will show that the twisted
self-duality equation $\ast \mathcal{G=SG}$, [1,2] when applied on
the Cartan form leads to the first-order formulation of the
equations of motion achieved in section two.

\section{The $D=8$ Matter Coupled Salam-Sezgin Supergravity}
The eight dimensional Salam-Sezgin supergravity with matter
coupling is constructed in [6]. N vector multiplets
$\{\lambda,A,\varphi^{i}\}$ ($\lambda$ is a Fermion and for
$i=1,2$ $\{\varphi^{i}\}$ are the scalars whereas $A$ is a
one-form field) are coupled to the original field content
$\{e_{\mu}^{m},\psi^{\mu},\chi,B_{\mu\nu},A_{i}^{\mu},\sigma\}$
where $\{e_{\mu}^{m}\}$ is the graviton, $\{\psi^{\mu},\chi\}$ are
the fermionic fields, $\{\sigma\}$ is the dilaton, $B_{\mu\nu}$ is
a two-form field and for $i=1,2$ $\{A_{i}^{\mu}\}$ are the
one-form fields. We will combine the original one-forms
$\{A_{i}^{\mu};i=1,2\}$ and the N vector multiplet one-forms
$\{A\}$ into a single set and we will denote them as $\{A^{j};
j=1,...,$N$+2\}$. Later we will also classify the 2N vector
multiplet scalars $\{\varphi^{i}\}$ as dilatons and axions as a
result of the solvable Lie algebra parametrization [8]. The single
dilaton $\{\sigma\}$ of the original theory and the 2N scalars
$\{\varphi^{i}\}$ of the vector multiplet which parameterize the
$SO(N,2)/SO(N)\times SO(2)$ K\"{a}hler coset manifold constitute
the scalar sector of the matter coupled theory. An explicit
representation parametrization for the scalar coset
$SO(N,2)/SO(N)\times SO(2)$ is given in [6] where the scalar
fields are not classified as dilatons and axions. Since $SO(N,2)$
is a non-compact real form of some semi-simple Lie group (for odd
N $so(N,2)$ is the non-compact real form of $B_{\frac{N+1}{2}}$
and for even N it is the non-compact real form of
$D_{\frac{N+2}{2}}$ and in general depending on N, $SO(N,2)$ is
not necessarily a split real form (maximally non-compact)
[7,9,10]) we will use the solvable Lie algebra parametrization
[8]. This parametrization introduces the dilatons which are
coupled to the generators of $\mathbf{h}_{k}$, the Lie algebra of
the maximal R-split torus of $SO(N,2)$ [9] and the axions which
are coupled to the positive root generators of $SO(N,2)$ which do
not commute with the elements of $\mathbf{h}_{k}$ [7,9]. Due to
the nature of the semi-simple, non-compact real form $SO(N,2)$ and
the fact that $SO(N)\times SO(2)$ is the maximal compact subgroup
of $SO(N,2)$ by using the Iwasawa decomposition [7] the coset
$SO(N,2)/SO(N)\times SO(2)$ can be parameterized as
\begin{equation}
\begin{aligned}
L&=\mathbf{g}_{H}\mathbf{g}_{N}\\
\\
&=e^{\frac{1}{2}\phi ^{i}(x)H_{i}}e^{\chi ^{m}(x)E_{m}}
\end{aligned}
\end{equation}
where $\{H_{i}\}$ for $i=1,...,$ dim$\mathbf{h}_{k}$ are the
generators of $\mathbf{h}_{k}$ and $\{E_{m}\}$ for
$m\in\Delta_{nc}^{+}$ are the positive root generators which
generate the orthogonal complement of $\mathbf{h}_{k}$ within the
solvable Lie algebra $\mathbf{s}_{0}$ of $SO(N,2)$, [7,8,9]. Also
the 2N scalars are now classified as $\{\phi^{i}\}$, the dilatons
for $i=1,...,$ dim$\mathbf{h}_{k}$ and $\{\chi^{m}\}$, the axions
for $m\in\Delta_{nc}^{+}$. In other words if we denote
dim$\mathbf{h}_{k}$ by $r$ and label the roots which are elements
of $\Delta_{nc}^{+}$ from $1$ to $n$ then we have $r+n=2$N. By
using the generalized transpose $\#$, [1,10] we can introduce the
internal metric $\mathcal{M}=L^{\#}L$. For $SO(N,2)$ since the
subgroup generated by the compact generators of the Lie algebra
$so(N,2)$ is an orthogonal group $\#$ coincides with the matrix
transpose in the fundamental representation [1]. Thus the
Lagrangian of the matter scalar sector of the eight dimensional
Salam-Sezgin supergravity [6] can be written as
\begin{equation}
 \mathcal{L}_{scalar}=\frac{1}{16}tr(d\mathcal{M}^{-1}\wedge \ast d\mathcal{M})
\end{equation}
where $\mathcal{M}=L^{T}L$. The bosonic Lagrangian of the $D=8$,
matter coupled Salam-Sezgin supergravity can now be given as [6]
\begin{equation}
\begin{aligned}
\mathcal{L}&=\frac{1}{4}R\ast1+\frac{3}{8} d\sigma\wedge \ast
d\sigma-\frac{1}{2}e^{2\sigma} G
\wedge \ast G\\
\\
&+\frac{1}{16}tr( d\mathcal{M}^{-1}\wedge \ast
d\mathcal{M})-\frac{1}{2}e^{\sigma} F\wedge\mathcal{M} \ast F
\end{aligned}
\end{equation}
where we have assumed the (N$+2$)-dimensional matrix
representation of $SO(N,2)$. We should also imply that the last
term in (2.3) above can be explicitly written as
$-\frac{1}{2}e^{\sigma}\mathcal{M}_{ij} F^{i}\wedge\ast F^{j}$
since $\mathcal{M}$ is a symmetric matrix having the components
$\mathcal{M}_{ij}=L_{i}^{a}L_{j}^{a}$ for $i,j,a=1,...,$N$+2$. The
(N$+2$) two-forms $\{F^{i}\}$ are the field strengths of
$\{A^{i}\}$, $F^{i}=dA^{i}$ and the Chern-Simons term $G$ is
defined as
\begin{equation}
 G=dB+\eta_{ij}F^{i}\wedge A^{j}.
\end{equation}
Here $B$ is the two-form field, the indices $\{i,j\}$ are running
from 1 to N$+2$ and $\eta$ is the metric corresponding to
$SO(N,2)$ explicitly $\eta=(-,-,+,+,+,...)$. We also have the
standard formulas
\begin{equation}
L^{T}\eta L=\eta\quad\quad,\quad\quad L^{-1}=\eta L\eta
\end{equation}
of the orthogonal matrix groups for $SO(N,2)$. The second identity
is due to the fact that the coset representatives $L$ can be
locally chosen as symmetric matrices $L^{T}=L$ which is evident
from the explicit parametrization given in [6] and the
transformation relating the parametrization of [6] and the
solvable Lie algebra parametrization which can be shown to exist
locally [11].

We will now give an identity for the Cartan generators $\{H_{i}\}$
which we will make use of in deriving the first-order equations
from the second-order ones. Firstly we should observe that
\begin{gather}
\partial_{i}L\equiv\frac{\partial L}{\partial\phi^{i}}=\frac{1}{2}H_{i}L\quad\quad,\quad\quad
\partial_{i}L^{T}\equiv\frac{\partial L^{T}}{\partial\phi^{i}}=\frac{1}{2}L^{T}H_{i}^{T},\notag\\
\notag\\
\partial_{i}L^{-1}\equiv\frac{\partial L^{-1}}{\partial\phi^{i}}=-\frac{1}{2}L^{-1}H_{i}\quad\quad,\quad\quad
\partial_{i}\mathcal{M}\equiv\frac{\partial\mathcal{M}}{\partial\phi^{i}}=\frac{1}{2}L^{T}(H_{i}+H_{i}^{T})L.
\end{gather}
By differentiating the equations in (2.5) with respect to
$\{\phi^{i}\}$ and by using the identities (2.5) and (2.6)
effectively we can show that $H_{i}^{T}\eta=-\eta H_{i}$ and
moreover
\begin{equation}
(H_{i}L)^{T}=H_{i}L
\end{equation}
for $i=1,...,$ dim$\mathbf{h}_{k}$. In obtaining the symmetry of
$H_{i}L$, namely the equation (2.7), we have implicitly used the
fact that the coset representatives can be chosen as symmetric
matrices. The identities (2.6) and (2.7) are matrix equations and
apparently they are valid in the (N+2)-dimensional representation
we have assumed. In order to find the field equations for the
non-gravitational fields we first vary the Lagrangian (2.3) with
respect to the fields $\sigma,B$ and $\{A^{i}\}$ and we find the
corresponding field equations as
\begin{gather}
\frac{3}{4}d(\ast d\sigma)=-e^{2\sigma} G\wedge \ast
G-\frac{1}{2}e^{\sigma}\mathcal{M}_{ij} F^{i}\wedge
\ast F^{j},\notag\\
\notag\\
d(e^{2\sigma}\ast G)=0,\notag\\
\notag\\
d(e^{\sigma}\mathcal{M}_{ij}\ast
F^{j})=-2e^{2\sigma}(\eta_{ij}F^{j}\wedge\ast G).
\end{gather}
One can follow the standard steps in [9,10] to find the
corresponding field equations for the scalars where the
formulation is done for a generic coset. Here we will not present
the complete derivation but we will only give the resulting
equations. The field equations for $\{\phi^{i}\}$ and
$\{\chi^{m}\}$ are
\begin{equation}
\begin{aligned}
d(e^{\frac{1}{2}\gamma _{i}\phi ^{i}}\ast U^{\gamma
})&=-\frac{1}{2}\gamma _{j}e^{\frac{1}{2}\gamma _{i}\phi
^{i}}d\phi ^{j}\wedge \ast U^{\gamma }\\
\\
&+\sum\limits_{\alpha -\beta =-\gamma }e^{\frac{1}{2} \alpha
_{i}\phi ^{i}}e^{\frac{1}{2}\beta _{i}\phi ^{i}}N_{\alpha ,-\beta
}U^{\alpha }\wedge \ast U^{\beta
},\\
\\
d(\ast d\phi ^{i})&=\frac{1}{2}\sum\limits_{\alpha\in\Delta_{nc}^{+}}^{}\alpha _{i}%
e^{\frac{1}{2}\alpha _{i}\phi ^{i}}U^{\alpha }\wedge e^{%
\frac{1}{2}\alpha _{i}\phi ^{i}}\ast U^{\alpha
}\\
\\
&-2e^{\sigma}((H_{i})_{n}^{a}L_{m}^{n}L_{j}^{a})\ast F^{m}\wedge
F^{j}
\end{aligned}
\end{equation}
where the indices associated with the dilatons and the Cartan
generators are from $1$ to dim$\mathbf{h}_{k}$ also
$\alpha,\beta,\gamma\in\Delta_{nc}^{+}$ [9]. The matrices
$\{(H_{i})_{n}^{a}\}$ are the ones corresponding to the generators
$\{H_{i}\}$ in the (N$+2$)-dimensional representation chosen. Our
formulation for the dilaton equation namely the second equation of
(2.9) is different from the one in [9] where the contribution from
the coupling of $\{A^{i}\}$ in the second equation of (2.9) was
expressed by using the weights of the representation. We keep the
original fields instead of their weight expansions. We have also
introduced
$U^{\alpha}=\mathbf{\Omega}_{\beta}^{\alpha}\,d\chi^{\beta}$ which
is not explicitly calculated in [9,10]. Here $\mathbf{\Omega}$ is
the matrix
\begin{equation}
\begin{aligned}
\mathbf{\Omega}&=\sum\limits_{n=0}^{\infty }\dfrac{\omega
^{n}}{(n+1)!}\\
\\
&=(e^{\omega}-I)\,\omega^{-1}
\end{aligned}
\end{equation}
with
$\omega_{\beta}^{\gamma}=\chi^{\alpha}K_{\alpha\beta}^{\gamma}$.
The structure constants $K_{\alpha\beta}^{\gamma}$ are defined as
$[E_{\alpha},E_{\beta}]=K_{\alpha\beta}^{\gamma}E_{\gamma}$ or
since $\{E_{\alpha}\}$ are the generators corresponding to a
subset, $\Delta_{nc}^{+}$ of the roots of $so(N,2)$ we have $[E_
{\alpha},E_{\beta}]=N_{\alpha,\beta}E_{\alpha+\beta}$. In other
words $K_{\beta\beta}^{\alpha}=0$,
$K_{\beta\gamma}^{\alpha}=N_{\beta,\gamma}$ if
$\beta+\gamma=\alpha$ and $K_{\beta\gamma}^{\alpha}=0$ if
$\beta+\gamma\neq\alpha$ in the root sense.

One can locally integrate the bosonic field equations (2.8) and
(2.9) by introducing dual fields and by using the fact that
locally a closed form is an exact one. By integration we mean to
extract an exterior derivative on both sides of the equations. It
is straightforward to write the first-order equations for (2.8).
On the other hand for the equations in (2.9) the first-order
formulation of the scalars must be treated separately due to their
non-linear nature. The integration of the terms in (2.9) which are
related to (2.2) can be done by an application of the dualisation
method of [1,2] only for the scalar sector separately that is the
method we have followed. If we introduce the dual four-form
$\widetilde{B}$, the set of five-forms $\{\widetilde{A}^{j}\}$ and
the six-forms
$\{d\widetilde{\sigma},d\widetilde{\phi}^{i},d\widetilde{\chi}^{m}\}$
we can locally derive the first-order equations as
\begin{gather}
e^{2\sigma}\ast G=d\widetilde{B},\notag\\
\notag\\
e^{\sigma}\mathcal{M}_{j}^{i}\ast F^{j}=-d\widetilde{A}^{i}+2d\widetilde{B}\wedge\eta_{j}^{i}A^{j},\notag\\
\notag\\
\ast d\sigma=d\widetilde{\sigma}-\frac{4}{3} B \wedge d\widetilde{B}+\frac{2}{3}\delta_{ij}A^{i}\wedge d\widetilde{A}^{j},\notag\\
\notag\\
e^{\frac{1}{2}\alpha_{i}\phi^{i}}(\mathbf{\Omega})_{l}^{\alpha}\ast
d\chi^{l}=(e^{\mathbf{\Gamma}}
e^{\mathbf{\Lambda}})_{j}^{\alpha+r}\widetilde{S}^{j},\notag\\
\notag\\
\frac{1}{2}\ast
d\phi^{m}=(e^{\mathbf{\Gamma}}e^{\mathbf{\Lambda}})_{j}^{m}\widetilde{S}^{j}+(H_{m})_{ji}A^{j}\wedge
d\widetilde{A}^{i}+\eta_{i}^{k} (H_{m})_{jk}A^{j}\wedge A^{i}
\wedge d \widetilde{B}
\end{gather}
where we have denoted $r\equiv\;$dim$\mathbf{h}_{k}$. The vector
$\overset{\rightharpoonup }{\widetilde{S}}$ is
$\widetilde{S}^{j}=\frac{1}{2}d\widetilde{\phi}^{j}$ for
$j=1,...,r$ and
$\widetilde{S}^{\alpha+r}=d\widetilde{\chi}^{\alpha}$ for
$\alpha\in\Delta_{nc}^{+}$. The matrices $\mathbf{\Gamma}$ and
$\mathbf{\Lambda}$ are introduced as $\mathbf{\Gamma }%
_{n}^{k}=\frac{1}{2}\phi ^{i}\,\widetilde{g}_{in}^{k}$ and
$\mathbf{\Lambda }_{n}^{k}=\chi ^{m}\widetilde{f}_{mn}^{k}. $The
coefficients $\{\widetilde{g}_{in}^{k}\}$ and
$\{\widetilde{f}_{mn}^{k}\}$ are as follows
\begin{gather}
\widetilde{f}_{\alpha m}^{n}=0,\quad\quad m\leq r\quad,\quad
\widetilde{f}_{\alpha ,\alpha +r}^{i}=\frac{1}{4}\alpha _{i},\quad\quad%
i\leq r\notag\\
\notag\\
\widetilde{f}_{\alpha ,\alpha +r}^{i}=0,\quad\quad i>r\quad,\quad
\widetilde{f}_{\alpha ,\beta +r}^{i}=0,\quad\quad i\leq r,\;%
\alpha \neq \beta \notag\\
\notag\\
\widetilde{f}_{\alpha ,\beta +r}^{\gamma +r}=N_{\alpha ,-\beta },\quad\quad \alpha -\beta =-\gamma,\;\alpha \neq \beta\notag\\
\notag\\
\widetilde{f}_{\alpha ,\beta +r}^{\gamma +r}=0,\quad\quad \alpha
-\beta \neq -\gamma ,\;\alpha \neq \beta.
\end{gather}

We should remind the reader of the fact that
$\alpha,\beta\in\Delta_{nc}^{+}$ and we assume that these roots
are enumerated from 1 to $2$N$-r$. The conditions on
$\alpha,\beta,\gamma$ in the last two lines of (2.12) must be
taken as root equations. In the next section we will see that
these coefficients are in fact the structure constants of the
commutators of the scalar and
 the dual scalar generators in other words $[E_{\alpha
},\widetilde{T}_{m}]=\widetilde{f}_{\alpha m}^{n}\widetilde{T}
_{n}$ , $[H_{i},\widetilde{T}_{m}]=\widetilde{g}_{im}^{n}
\widetilde{T}_{n}$ where $\widetilde{T}_{i}=\widetilde{H}_{i}$ for
$i=1,...,r$ and $\widetilde{T}_{\alpha+r}=\widetilde{E}_{\alpha}$
for $\alpha\in\Delta_{nc}^{+}$. Here $\{\widetilde{H}_{i}\}$ are
the generators which we will assign to the fields
$\{\widetilde{\phi}^{i}\}$ and $\{\widetilde{E}_{\alpha}\}$ are
the generators which will be assigned to the fields
$\{\widetilde{\chi}^{\alpha}\}$. The last term in the last
equation of (2.11) needs attention since if its exterior
derivative is taken, in order to obtain the corresponding term in
the second-order equations, one needs to make use of the fact that
as, $H_{i}^{T}\eta=-\eta H_{i}$ and since $\eta$ is a diagonal
matrix, $H_{i}\eta$ must have anti-symmetric matrix
representatives. One also needs to make use of the identities
(2.7) to reach the corresponding second-order equation in (2.9)
when the last equation of (2.11) is differentiated. We are using
the Euclidean metric to raise and lower the indices whenever
necessary for notational purposes.

These first-order equations will be re-formulated as a twisted
self-duality equation after we introduce the dualisation of the
theory resulting in a coset model in the next section.

\section{The Coset Formulation}

In this section by following the outline of [1,2] we will
generalize the coset formulation of the scalar manifold
$SO(N,2)/SO(N)\times SO(2)$ to the entire bosonic sector of the
$D=8$ matter coupled Salam-Sezgin supergravity. Our aim is to
construct a coset representative which will be used to realize the
original field equations. The group which is generated by the
generators we introduce to construct the coset representative
within the dualisation, in general becomes the symmetry group of
the Cartan form ${\mathcal{G}}$ which is generated by the coset
representative map. We need to construct a new algebra which will
lead to the non-linear realization of the bosonic field equations.
The first-order equations (2.11) derived in the last section will
be formulated as a twisted self-duality equation
$\ast\mathcal{G=SG}$ where $\mathcal{S}$ is a pseudo-involution of
the dualized Lie superalgebra. The full symmetry group of the
first-order equations is in general bigger than the symmetry group
of the Cartan form.

We will assign a generator for each field. The untilded original
generators are $\{K,V_{i},Y,H_{j},E_{m}\}$ for the fields
$\{\sigma,A^{i},B,\phi^{j},\chi^{m}\}$ respectively. Their duals
are introduced as
$\{\widetilde{K},\widetilde{V}_{i},\widetilde{Y},
\widetilde{H}_{j},\widetilde{E}_{m}\}$ for the dual fields
$\{\widetilde{\sigma},\widetilde{A}^{i},\widetilde{B},
\widetilde{\phi}^{j},\widetilde{\chi}^{m}\}$. We require that the
Lie superalgebra to be constructed from the generators has the
$\mathbb{Z}_{2}$ grading as usual. For this reason the generators
will be chosen as odd if the corresponding potential is an odd
degree differential form and otherwise even. In particular
$\{V_{i},\widetilde{V}_{i}\}$ are odd generators and the rest of
the generators are even. We will also consider a differential
graded algebra generated by the differential forms and the
generators which are coupled to the fields. This algebra covers
the Lie superalgebra of the field generators. Therefore the odd
(even) generators behave like odd (even) degree differential forms
under this graded differential algebra structure when they commute
with the exterior product. The odd generators obey
anti-commutation relations while the even ones and the mixed
couples obey commutation relations.

As it will be clear later the structure constants of this new
algebra will be chosen so that they will lead to the correct
second-order equations (2.8) and (2.9). Firstly let us consider
the map
\begin{equation}
\nu=e^{\frac{1}{2}\phi^{j}H_{j}}e^{\chi^{m}E_{m}}e^{\sigma
K}e^{A^{i}V_{i}}e^{\frac{1}{2}BY}.
\end{equation}
The corresponding Cartan form $\mathcal{G}=d\nu\nu^{-1}$ can be
expanded in terms of the original generators and it satisfies the
Cartan-Maurer equation
$d\mathcal{G}-\mathcal{G}\wedge\mathcal{G}=0$ owing to its
definition. The coset representative by including the dual
generators as well will be chosen as
\begin{equation}
\nu^{\prime}=e^{\frac{1}{2}\phi^{j}H_{j}}e^{\chi^{m}E_{m}}e^{\sigma
K}e^{A^{i}V_{i}}e^{\frac{1}{2}BY}
e^{\frac{1}{2}\widetilde{B}\widetilde{Y}}e^{\widetilde{A}^{i}\widetilde{V}_{i}}e^{\widetilde{\sigma}
\widetilde{K}}e^{\widetilde{\chi}^{m}\widetilde{E}_{m}}e^{\frac{1}{2}\widetilde{\phi}^{j}\widetilde{H}_{j}}.
\end{equation}
The Cartan form $\mathcal{G}^{\prime}=d\nu^{\prime}\nu^{\prime-1}$
can also be calculated in the expansion of the original and the
dual generators. For the calculation of both of the Cartan forms
$\mathcal{G}$ and $\mathcal{G}^{\prime}$ we need to know the
algebra structure that the generators obey. This structure will be
constructed in a way that the twisted self-duality equation
$\ast\mathcal{G}^{\prime}=\mathcal{SG}^{\prime}$ will lead to the
correct first-order equations. The action of the pseudo-involution
$\mathcal{S}$ on the generators is as follows
\begin{gather}
\mathcal{S}Y=\widetilde{Y}\quad,\quad\mathcal{S}K=\widetilde{K}\quad,\quad
\mathcal{S}E_{m}=\widetilde{E}_{m}\quad,\quad\mathcal{S}H_{i}=\widetilde{H}_{i},\notag\\
\notag\\
\mathcal{S}\widetilde{Y}=Y\quad,\quad\mathcal{S}\widetilde{K}=K\quad,\quad
\mathcal{S}\widetilde{E}_{m}=E_{m}\quad,\quad\mathcal{S}\widetilde{H}_{i}=H_{i},\notag\\
\notag\\
\mathcal{S}V_{i}=\widetilde{V}_{i}\quad,\quad\mathcal{S}\widetilde{V}_{i}=-V_{i}.
\end{gather}
The Cartan form $\mathcal{G}^{\prime}=d\nu^{\prime}\nu^{\prime-1}$
also obeys the Cartan-Maurer equation
\begin{equation}
d\mathcal{G}^{\prime}-\mathcal{G}^{\prime}\wedge\mathcal{G}^{\prime}=0
\end{equation}
One can primarily use the twisted self-duality condition
$\ast\mathcal{G}^{\prime}=\mathcal{SG}^{\prime}$ to construct the
generator expansion of $\mathcal{G}^{\prime}$ and obtain an
expression only in terms of the original fields then calculate
(3.4). This equation must lead to the second-order field equations
(2.8) and (2.9) [2]. In the calculation of the generator expansion
of the Cartan form $\mathcal{G}^{\prime}$ we firstly calculate
$\mathcal{G}$ in terms of the unknown structure constants of the
original field generators. $\mathcal{G}$ constitutes the part of
$\mathcal{G}^{\prime}$ which is composed of the original
generators. We can use the twisted self-duality condition we
propose on $\mathcal{G}^{\prime}$ to generate the other part of
$\mathcal{G}^{\prime}$ which is composed of the dual generators.
This is possible since when one inspects the Lie superalgebra
structure in general, in order to obtain the correct field
equations one finds that the commutation or the anti-commutation
relations of the original generators must lead to the original
generators and a pair of an original and a dual generator would
lead to a dual generator under the algebra product while the
product of two dual generators would vanish. When one calculates
$\mathcal{G}^{\prime}$ as explained above and then insert it in
the Cartan-Maurer equation, the result can be compared with (2.8)
and (2.9) to read the desired commutation and the anti-commutation
relations of the Lie superalgebra of the field generators. We will
not give the details of this long calculation but only present the
results. The resulting commutation and the anti-commutation
relations of the original and the dual generators apart from the
purely scalar commutators are
\begin{gather}
[K,V_{i}]=\frac{1}{2}V_{i}\quad,\quad[K,Y]=Y\quad,\quad[K,\widetilde{Y}]=-\widetilde{Y},\notag\\
\notag\\
[\widetilde{V}_{k},K]=\frac{1}{2}\widetilde{V}_{k}\quad,\quad\{V_{i},V_{j}\}=\eta_{ij}Y\quad,\quad
[H_{l},V_{i}]=(H_{l})_{i}^{k}V_{k}\notag\\
\notag\\
[E_{m},V_{i}]=(E_{m})_{i}^{j}V_{j}\quad,\quad\{V_{l},\widetilde{V}_{k}\}=\frac{2}{3}\delta_{lk}\widetilde{K}
+\underset{i}{\sum
}(H_{i})_{lk}\widetilde{H}_{i},\notag\\
\notag\\
[V_{k},\widetilde{Y}]=-4\,\eta_{k}^{l}\,\widetilde{V}_{l}\quad,\quad[Y,\widetilde{Y}]=-\frac{16}{3}\widetilde{K}
\quad,\quad[H_{i},\widetilde{V}_{k}]=-(H_{i}^{T})_{k}^{m}\widetilde{V}_{m},\notag\\
\notag\\
[E_{\alpha},\widetilde{V}_{k}]=-(E_{\alpha}^{T})_{k}^{m}\widetilde{V}_{m}.
\end{gather}
The matrices ($(H_{m})_{i}^{j}$, $(E_{\alpha})_{i}^{j}$) are the
images of the corresponding generators ($H_{m},E_{\alpha}$) under
the representation chosen respectively. Also the matrices
($(H_{m}^{T})_{i}^{j}$, $(E_{\alpha}^{T})_{i}^{j}$) are the matrix
transpose of ($(H_{m})_{i}^{j}$, $(E_{\alpha})_{i}^{j}$). The
scalar generators and the generators coupled to the 6-form dual
fields of the scalars namely
$\{H_{i},E_{m},\widetilde{E}_{m},\widetilde{H}_{i}\}$ constitute a
subalgebra with the following commutators
\begin{gather}
[H_{j},E_{\alpha }]=\alpha _{j}E_{\alpha }\quad ,\quad [E_{\alpha
},E_{\beta }]=N_{\alpha ,\beta }E_{\alpha+\beta},\notag\\
\notag\\
[H_{j},\widetilde{E}_{\alpha }]=-\alpha _{j}\widetilde{E}_{\alpha }\quad ,\quad [%
E_{\alpha },\widetilde{E}_{\alpha }]=\frac{1}{4}\overset{r}{\underset{j=1}{%
\sum }}\alpha _{j}\widetilde{H}_{j},\notag\\
\notag\\
[E_{\alpha },\widetilde{E}_{\beta }]=N_{\alpha ,-\beta }\widetilde{E}%
_{\gamma },\quad\quad\alpha -\beta =-\gamma,\alpha \neq \beta,
\end{gather}
where $i,j=1,...,r$ and $\alpha,\beta,\gamma\in\Delta_{nc}^{+}$.
Here $\Delta_{nc}^{+}$ is a subset of the positive roots of
$SO(N,2)$ whose corresponding generators do not commute with the
elements of $\mathbf{h}_{k}$ in $so(N,2)$ [9]. The remaining
commutators and the anti-commutators which are not listed in (3.5)
and (3.6) vanish indeed. We observe that the structure constants
of the commutators of the scalar and the dual scalar generators in
(3.6) are the coefficients introduced in (2.12). We can now
calculate the doubled field strength
$\mathcal{G}^{\prime}=d\nu^{\prime}\nu^{\prime-1}$ explicitly by
using the above commutation and the anti-commutation relations.
From the definition of the coset element in (3.2) by using (3.5)
and (3.6) the calculation of
$\mathcal{G}^{\prime}=d\nu^{\prime}\nu^{\prime-1}$ yields
\begin{equation}
\begin{aligned}
\mathcal{G}^{\prime}&=\frac{1}{2}d\phi^{i}H_{i}+e^{\frac{1}{2}\alpha_{i}\phi^{i}}U^{\alpha}E_{\alpha}
+d\sigma K+e^{\frac{1}{2}\sigma}L_{i}^{k}dA^{i}V_{k}+\frac{1}{2}e^{\sigma}GY\\
\\
&+\frac{1}{2}e^{-\sigma}d\widetilde{B}\widetilde{Y}+(-\frac{4}{3}B\wedge
d\widetilde{B}+ \frac{2}{3}A^{j}\wedge
d\widetilde{A}^{i}\delta_{ij}+d\widetilde{\sigma})\widetilde{K}+\overset{\rightharpoonup
}{\widetilde{\mathbf{T}}}(\overset{\rightharpoonup
}{\mathbf{J}}+e^{\mathbf{\Gamma}}e^{\mathbf{\Lambda}}\overset{\rightharpoonup
}{\widetilde{\mathbf{S}}})\\
\\
&+(e^{-\frac{1}{2}\sigma}((L^{T})^{-1})_{k}^{l}d\widetilde{A}^{k}+2e^{-\frac{1}{2}\sigma}((L^{T})^{-1})_{k}^{l}
\,\eta_{i}^{k}\,A^{i}\wedge d\widetilde{B})\widetilde{V}_{l}.
\end{aligned}
\end{equation}
In the derivation of (3.7) we have used the matrix identities
$de^{X}e^{-X}=dX+\frac{1}{2!}[X,dX]+\frac{1}{3!}[X,[X, dX]]+....$
and $e^{X}Ye^{-X}=Y+[X,Y]+\frac{1}{2!}[X,[X,Y]]+....$ . We have
defined the set $\{U^{\alpha}\}$ in section two. Although $L$ is a
symmetric matrix we keep $L^{T}$ in the above expression for an
easier comparison of the first-order equations we will obtain from
(3.7) with the ones previously calculated in (2.11). Here the row
vector $\overset{\rightharpoonup }{\widetilde{\mathbf{T}}}$ is
defined as $\widetilde{\mathbf{T}}_{i}=\widetilde{H}_{i}$ for
$i=1,...,r$ and
$\widetilde{\mathbf{T}}_{\alpha+r}=\widetilde{E}_{\alpha}$ for
$\alpha\in\Delta_{nc}^{+}$. As defined in section two the column
vector $\overset{\rightharpoonup }{\widetilde{\mathbf{S}}}$ is
$\widetilde{\mathbf{S}}^{i}=\frac{1}{2}d\widetilde{\phi}^{i}$ for
$i=1,...,r$ and
$\widetilde{\mathbf{S}}^{\alpha+r}=d\widetilde{\chi}^{\alpha}$ for
$\alpha\in\Delta_{nc}^{+}$ (we have already assumed that the roots
in $\Delta_{nc}^{+}$ are labelled by integer indices from $1$ to
$n$ therefore $r+n=2$N). The $\overset{\rightharpoonup
}{\mathbf{J}}$ term in (3.7) is a result of the coupling between
the scalars and the one-form potentials in (2.3) and we define it
as
\begin{gather}
\mathbf{J}^{m}=(H_{m})_{ji}A^{j}\wedge
 d\widetilde{A}^{i}+\eta_{i}^{k}
(H_{m})_{jk}A^{j}\wedge A^{i} \wedge d \widetilde{B}\quad,\quad
m=1,...,r\notag\\
\notag\\
\mathbf{J}^{\alpha+r}=0\quad,\quad\alpha\in\Delta_{nc}^{+}.
\end{gather}
In (3.7) more explicitly we have
\begin{equation}
\begin{aligned}
\overset{\rightharpoonup
}{\widetilde{\mathbf{T}}}(\overset{\rightharpoonup
}{\mathbf{J}}+e^{\mathbf{\Gamma}}e^{\mathbf{\Lambda}}\overset{\rightharpoonup
}{\widetilde{\mathbf{S}}})&=\overset{r}{\underset{m=1}{\sum}}((e^{\mathbf{\Gamma}}e^{\mathbf{\Lambda}})_{j}^{m}
\widetilde{\mathbf{S}}^{j}+(H_{m})_{ji}A^{j}\wedge
d\widetilde{A}^{i}\\
\\
&+\eta_{i}^{k} (H_{m})_{jk}A^{j}\wedge A^{i} \wedge
d\widetilde{B})\widetilde{H}_{m}+\underset{\alpha\in\Delta_{nc}^{+}}{\sum}(e^{\mathbf{\Gamma}}e^{\mathbf{\Lambda}})_{j}^{\alpha+r}
\widetilde{\mathbf{S}}^{j}\widetilde{E}_{\alpha}.
\end{aligned}
\end{equation}
The next step is to show that if we apply the twisted self-duality
equation $\ast\mathcal{G}^{\prime}=\mathcal{SG}^{\prime}$ on (3.7)
by using (3.3) we get the correct first-order equations (2.11).
Thus the twisted self-duality equation
$\ast\mathcal{G}^{\prime}=\mathcal{SG}^{\prime}$ gives
\begin{gather}
\frac{1}{2}e^{\sigma}\ast G=\frac{1}{2}e^{-\sigma}d\widetilde{B},\notag\\
\notag\\
e^{\frac{1}{2}\sigma}\L_{j}^{i}\ast
dA^{j}=-e^{-\frac{1}{2}\sigma}((L^{T})^{-1})_{j}^{i}d\widetilde{A}^{j}+2e^{-\frac{1}{2}\sigma}
((L^{T})^{-1})_{j}^{i}\,\eta_{k}^{j}\, d\widetilde{B}\wedge A^{k},\notag\\
\notag\\
\ast d\sigma=d\widetilde{\sigma}-\frac{4}{3}B\wedge d\widetilde{B}+\frac{2}{3}\delta_{ij}A^{j}\wedge d\widetilde{A}^{i},\notag\\
\notag\\
e^{\frac{1}{2}\alpha_{i}\phi^{i}}(\mathbf{\Omega})_{l}^{\alpha}\ast
d\chi^{l}=(e^{\mathbf{\Gamma}}
e^{\mathbf{\Lambda}})_{j}^{\alpha+r}\widetilde{\mathbf{S}}^{j},\notag\\
\notag\\
\frac{1}{2}\ast
d\phi^{m}=(e^{\mathbf{\Gamma}}e^{\mathbf{\Lambda}})_{j}^{m}\widetilde{\mathbf{S}}^{j}+
(H_{m})_{ji}A^{j}\wedge d\widetilde{A}^{i}+\eta_{i}^{k}
(H_{m})_{jk}A^{j}\wedge A^{i} \wedge d \widetilde{B}.
\end{gather}
These equations are the same with the first-order equations (2.11)
which are obtained by directly integrating the second-order
equations (2.8) and (2.9). We should also state that in (2.11) the
first-order formulation of the scalar sector (2.2) excluding the
matter coupling is done implicitly and separately by using the
dualisation method only on the scalars. The second equation in
(3.10) may seem to be different however it is the second equation
of (2.11) multiplied by $(L^{T})^{-1}$ from the left. This result
also separately justifies the proper choice of the commutation and
the anti-commutation relations in (3.5) and (3.6).

\section{Conclusion}
We have given the non-linear realization of the Salam-Sezgin $D=8$
supergravity which is coupled to N vector multiplets [6]. After
obtaining the second and the first-order equations of motion in
section two we have doubled the non-gravitational bosonic field
content by defining dual fields and constructed a Lie superalgebra
which leads to a coset formulation of the bosonic sector of the
$D=8$ matter coupled Salam-Sezgin supergravity in section three.

This work is an example of the dualisation method [1,2] which was
used to formulate the maximal supergravities in $D\leq11$ as
non-linear coset models. The formulation presented here shows that
the bosonic sectors of the matter coupled supergravities also
exhibit non-linear sigma model structures with a symmetry group
which covers the rigid symmetry group of the scalar Lagrangian.
The symmetry group of the doubled field strength is generated by
the Lie superalgebra introduced in section three, however, the
symmetry group of the twisted self-duality equation in other words
the first-order equations is still to be determined.

The non-linear realization of the bosonic sector may be enlarged
to include the gravity as well. A full dualisation of the gravity
and the other bosonic fields would lead to a larger symmetry
group. This group would cover the Lorentz group and the Lie
supergroup constructed here. The complete non-linear realization
of the bosonic sector may then be performed by taking the Lorentz
group as the local symmetry group also by considering the
simultaneous non-linear realization of the conformal group.
Furthermore the method presented in [3,4,5] may be applied to
detect the Kac-Moody symmetries of the matter coupled Salam-Sezgin
$D=8$ supergravity.


\begin{thebibliography}{99}
\bibitem{}
 E. Cremmer, B. Julia, H. L\"{u} and C. N. Pope, ``\textit{Dualisation of
 dualities}", Nucl. Phys. \textbf{B523} (1998) 73, hep-th/9710119.
\bibitem{}
 E. Cremmer, B. Julia, H. L\"{u} and C. N. Pope, ``\textit{Dualisation of dualities II : Twisted self-duality of doubled fields and superdualities}",
Nucl. Phys. \textbf{B535} (1998) 242, hep-th/9806106.
\bibitem{}
 P. West, ``\textit{Hidden superconformal symmetry in M theory}", JHEP
\textbf{08} (2000) 007, hep-th/0005270.
\bibitem{}
 P. West, ``\textit{E(11) and M theory}", Class. Quant. Grav. \textbf{18} (2001) 4443, hep-th/0104081.
\bibitem{}
 I. Schnakenburg and P. West, ``\textit{Kac-Moody symmetries of IIB
 supergravity}", Phys. Lett. \textbf{B517} (2001) 421, hep-th/0107181.
\bibitem{}
 A. Salam and E. Sezgin, ``\textit{d=8 Supergravity : Matter couplings, gauging and Minkowski
 compactification}", Phys. Lett. \textbf{B154} (1985) 37.
\bibitem{}
 S. Helgason, ``\textbf{Differential Geometry, Lie Groups and Symmetric
 Spaces}", (Graduate Studies in Mathematics 34, American Mathematical Society
 Providence R. I. 2001).
\bibitem{}
L. Andrianopoli, R. D'Auria, S. Ferrara, P. Fre and M. Trigiante,
``\textit{R-R Scalars, U-duality and solvable Lie algebras}",
Nucl. Phys. \textbf{B496} (1997) 617, hep-th/9611014.
\bibitem{}
 A. Keurentjes, ``\textit{The group theory of oxidation II : Cosets of non-split
 groups}", Nucl. Phys. \textbf{B658} (2003) 348, hep-th/0212024.
\bibitem{}
 A. Keurentjes, ``\textit{The group theory of oxidation}", Nucl. Phys.
\textbf{B658} (2003) 303, hep-th/0210178.
\bibitem{}
D. H. Sattinger and O. L. Weaver, ``\textbf{Lie Groups and
Algebras with
 Applications to Physics, Geometry and Mechanics}", (Springer-Verlag
 New York Inc. 1986).












\end{thebibliography}
\end{document}